\documentclass[12pt]{iopart}
\usepackage{iopams}
\usepackage{graphicx}
\usepackage{cite}

\begin{document}

\title[Quantum mechanical scattering on time-dependent potentials: TDNEGF method]{Quantum mechanical scattering on time-dependent potentials using nonequilibrium Green's functions}
\author{P\'{e}ter F\"{o}ldi}
\address{Department of Theoretical Physics, University of Szeged, Tisza Lajos k\"{o}r%
\'{u}t 84-86, H-6720 Szeged, Hungary}
\address{ELI-ALPS, ELI-HU Non-profit Ltd., Dugonics t\'{e}r 13, H-6720 Szeged, Hungary}

\begin{abstract}
Time-dependent nonequilibrium Green's functions (TDNEGF) are shown to provide a flexible, effective tool for the description of quantum mechanical single particle scattering on a spatially localized, time-dependent potential. Focusing on numerical methods, arbitrary space and time dependence of the potential can be treated, provided it is zero before an initial time instant. In this case, appropriate version of the Dyson and Keldysh equations lead to a transparent description with clear physical interpretation. The interaction of a short laser pulse and an electron propagating initially in free space is discussed as an example.
\end{abstract}

\maketitle

\section{Introduction}

Time-dependent scattering problems appear e.g.~in atomic, molecular and solid state physics, but they are of relevance also in nanoscale electronic devices driven by an alternating or pulse-like bias. In spite of the difference between these physical systems, the mathematical models that can describe these phenomena are similar. In the current paper we focus on methods based on nonequlibrium Green's functions, which -- besides providing an adequate description of many particle quantum systems -- have important applications also in transport-related problems in solids \cite{D95}. Ballistic (coherent) transport in the single electron approximation (which is often used for nanometer-sized samples) is equivalent to single particle quantum mechanical scattering problems. That is, methods developed for the description of transport processes in solid state systems have a more general scope in scattering theory.

The formalism based on time-dependent, nonequilibrium Green's functions (TDNEGFs) has been initiated in the early 1960s \cite{KB62,K65} but has many fundamental aspects that are discussed also in recent textbooks on field theory \cite{R11,K11}. This approach has been successfully applied to transport phenomena mainly in nanoscale solid state devices and static potentials (for a recent summary, see \cite{D10}). TDNGF-based description of time-dependent transport in mesoscopic systems was considered already in 1993 \cite{WJM93}, but numerically effective methods have been developed more recently. The details of these methods can be found in reference \cite{G14}, where the connection between TDNEGF and other approaches \cite{C80,BB00} is also discussed.

In the current paper we investigate how the formalism based on TDNEGFs can be used to describe scattering phenomena. For the sake of simplicity, we consider one spatial dimension, which is useful for the clear interpretation of the final formulas, but does not mean a necessary restriction. (Generalization to two and three dimensions can be done without essential difficulties.) We consider a monoenergetic plane wave that propagates initially in free space, i.e., the potential is zero. Then, at $t=0,$ a localized potential emerges, causing e.g.~time-dependent reflection phenomena. In order to keep the generality of the treatment, a spatial grid is applied, i.e., the position variable is discretized. Note that this is the only numerical approximation in the model (allowing calculations for potentials with arbitrary space and time dependence), all other calculations leading to the dynamical equations are analytic.

Time-dependent theory of nonequilibrium Green's functions rely on the Dyson and Keldysh equations as central concepts, and usually the main goal is to determine the so-called lesser Green's function from which direct physical consequences can be drawn. The solution of the equations resulting directly from the TDNEGF theory mean, however, generally an extremely complex task, the numerical costs of such calculations are tremendous. In our case, as it is going to be shown, the combination of the Dyson and Keldysh equations leads to a time evolution that can be calculated efficiently, and have a clear physical interpretation as well. Namely, the dynamics has to be solved only in the spatial region where the potential is nonzero, in this domain the usual time-dependent Schr\"{o}dinger equation is valid, and TDNEGF theory takes the boundary conditions into account by allowing the wave function to leave the "region of interest" without disturbance. These "perfectly transparent" boundary conditions turn the Schr\"{o}dinger equation into an integro-differential equation, but the relevant integrals involve only the boundary points of the finite region in which the dynamics is to be calculated. Note that these results are analogous to those derived in reference \cite{G14} in a different way for two dimensional ballistic transport (and applied for the description of "flying qubits" in coupled quantum wires), but here the focus is on general scattering phenomena and the interpretation of the dynamical equations. Additionally, as an application, in the last section we calculate how a short laser pulse interacting with a monoenergetic electron beam can create electron density fluctuations.

\section{Statement of the problem}
\label{problemsec}
\begin{figure}[tbh]
\centering
\includegraphics[width=12cm]{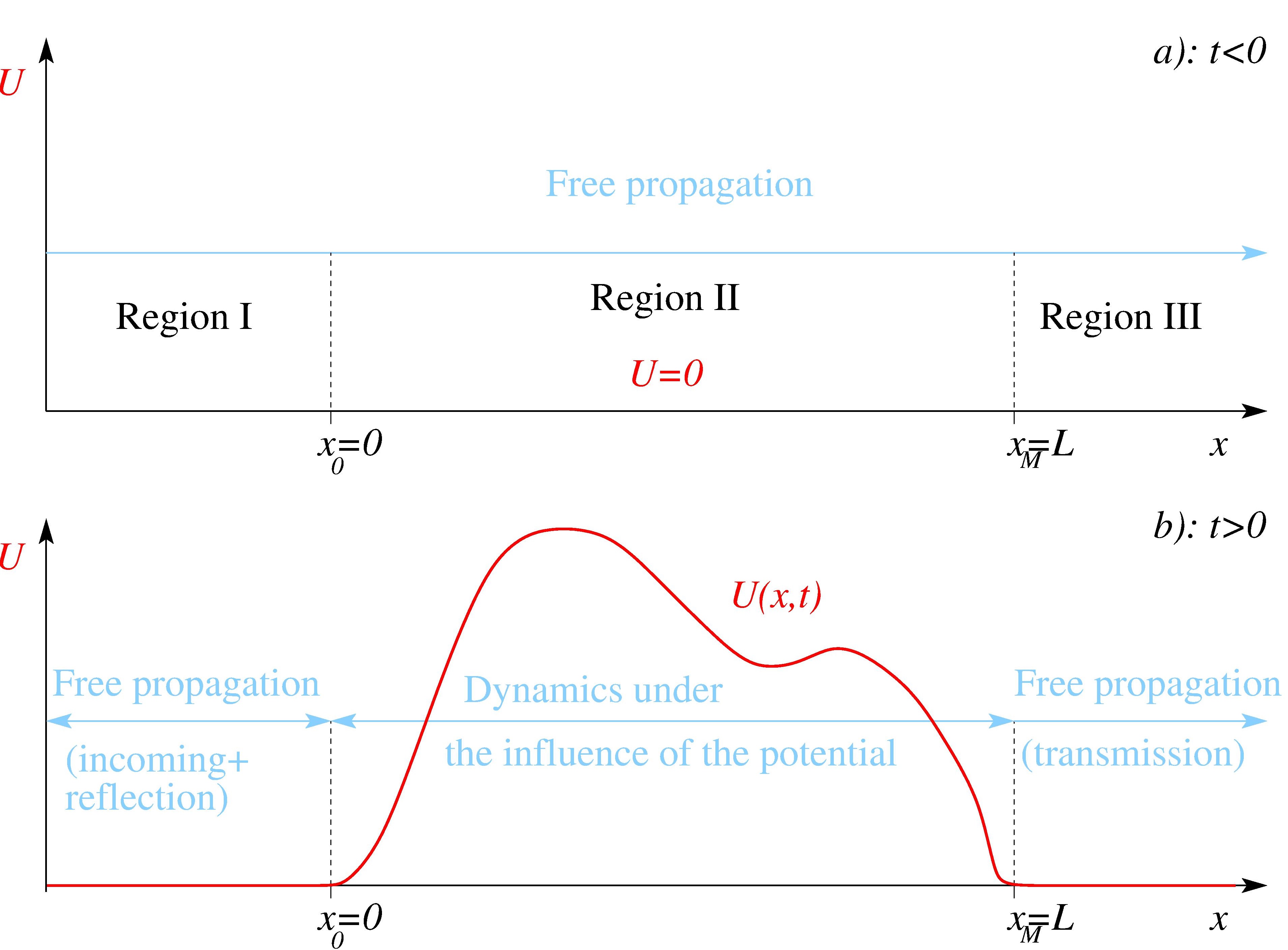}
\caption{(Color online) Schematic representation of the problem we consider.}
\label{problemfig}
\end{figure}

Let us consider the problem shown schematically in figure \ref{problemfig}. For $t<0,$ we have a monoenergetic quantum mechanical plane wave propagating in free space
\begin{equation}
\Phi(x,t<0)=e^{i\left[k(E)x-{Et}\right]}.
\label{wf}
\end{equation}
Here, we set $\hbar=1.$ (This convention will be used throughout the paper.) According to the Schr\"{o}dinger equation,
\begin{equation}
k(E)=\sqrt{2mE},
\label{invdispersion}
\end{equation}
where $m$ denotes the mass of the particle, and the direction of propagation has been chosen to correspond to figure \ref{problemfig}. Then, we assume that a localized potential, the support of which is the interval $[0,L],$ is turned on at $t=0.$  That is,
 \begin{equation}
U(x,t)=0, \ \ \mathrm{if}\  t<0, \ \mathrm{or}\  x\notin [0,L],
\label{potential}
\end{equation}
otherwise the dependence of $U$ on space and time can be practically arbitrary.

In the following we present a solution to the problem described above. Clearly, it is not the only possible approach, but the TDNEGF-based method treats the scattering-like  boundary conditions in a specific, exact way. In 1D, less sophisticated brute force numerical methods can solve the problem, but they are difficult to transfer to two or more dimensions (which is quite straightforward using TDNEGF \cite{G14}).

\section{Green's functions and discretization}
In single particle quantum mechanics, we can call a function with two spacetime arguments Green's function, if it is the "inverse of the Schr\"odinger operator," or, in other words, it is the fundamental solution of the time-dependent Schr\"odinger equation. That is,
\begin{equation}
\left(i\frac{\partial}{\partial t_1} -H(x_1, t_1) \right)G(x_1,t_1, x_2, t_2)=\delta(t_1-t_2)\delta(x_1-x_2),
\end{equation}
which, however, does not define $G(x_1,t_1, x_2, t_2)$ uniquely. Probably the most important member of this class of Green's function is the retarded propagator, which satisfies the additional condition
\begin{equation}
G^R(x_1,t_1, x_2, t_2)=0 \ \ \mathrm{for} \ \  t_1<t_2.
\label{GRzero}
\end{equation}
As it can be shown \cite{R11}, $G^R$ indeed propagates the wave function of the system forward in time, which follows from the fact that it essentially equals to the position basis matrix elements of the time evolution operator:
\begin{equation}
G^R(x_1,t_1, x_2, t_2)=-i\langle x_1|U(t_1, t_2)|x_2\rangle \ \ \mathrm{for} \ \  t_1<t_2.
\end{equation}

As we shall see later, it is useful to embed these functions in a wider context, and consider the one particle subspace of a many particle Fock-space and express $G^R$ in terms of the expectation value of field operator products:
\begin{equation}
G^R(1,2)=G^R\left(x_1,t_1, x_2,t_2\right)=-i\Theta(t_1-t_2)\left\langle\left[\Psi(x_1,t_1),\Psi^\dagger(x_2,t_2)\right]\right\rangle. \\
\end{equation}
The commutator above generally corresponds to the case of fermions, but we do not need to distinguish bosons and fermions when considering a single particle. The expectation value has to be calculated in the Heisenberg picture ("static") state of the system, which is generally represented by a statistical operator, but in our case it simply corresponds to the single particle state (\ref{wf}). The Heaviside function $\Theta$ ensures that the retarded wave function disappears for $t_1<t_2.$

The theory of (many particle) nonequilibrium Green's function also uses the lesser Green's function (two point correlation function), which can be written as
\begin{equation}
G^<(1,2)=G^<\left(x_1,t_1, x_2,t_2\right)=i\left\langle\Psi^\dagger(x_2,t_2)\Psi(x_1,t_1)\right\rangle.
\end{equation}
These functions can be used to calculate physical quantities, e.g., the local particle density reads
\begin{equation}
\label{dens}
n(x,t)=-iG^<\left(x,t, x,t\right).
\end{equation}
[Note that the density above is not normalized in the sense that its integral over the whole infinite spatial domain is not unity -- for the boundary conditions that we consider this kind of normalization is not even possible. Instead, we chose the convention $n(x,t)=1$ for the monoenergetic plane wave given by equation (\ref{wf}).]

In the following we are going to show how many particle TDNEGF dynamical equations (that describe the time evolution of $G^R$ and $G^<$) give rise to physically transparent equations in the one-particle subspace leading to an appropriate method to treat time-dependent scattering problems.

\bigskip

The equations of motion for $G^R$ and $G^<$ can be obtained using a method developed by Keldysh and others \cite{KB62,K65}. Since our aim is to focus on the general numerical solution of these equations, we recall their discretized version here, i.e., we consider a spatial grid. Let the stepsize be denoted by $a,$ i.e., we have $x_j=ja$ as grid points ($j=0,\ldots,M$ for region II in figure \ref{problemfig}). Then the Green's functions turn into matrices
\begin{eqnarray}
G^R_{ij}\left(t_1,t_2\right)&=&G^R\left(x_i,t_1, x_j,t_2\right), \\
G^<_{ij}\left(t_1,t_2\right)&=&G^<_{ij}\left(x_i,t_1, x_j,t_2\right),
\end{eqnarray}
and the input plane wave (\ref{wf}) also has to be replaced by its discretized version
\begin{equation}
\Phi_j(t<0)=\Phi(x_j,t<0)=e^{i\left[k(E)ja-Et\right]}.
\label{wf2}
\end{equation}
For the sake of simplicity, the Laplacian appearing in the one-particle Hamiltonian
\begin{equation}
H=H^0 + U(x,t)=-\frac{1}{2m} \ \Delta + U(x,t),
\label{H0}
\end{equation}
is usually represented by a simple three-point finite difference operator, leading to
\begin{equation}
\label{Ham}
H_{i,j}(t)=H^0_{i,j}+ U_{i,j}(t)=\left\{\begin{array}{ll}
      U(x_i,t)+\frac{1}{ma^2}, & \mathrm{if} \ i-j=0, \\
      -\frac{1}{2ma^2}, & \mathrm{if} \ |i-j|=1,  \\
      0 & \mathrm{otherwise}.
\end{array}
\right.
\end{equation}
The discrete version of the "dispersion relation" corresponding to this Hamiltonian reads
\begin{equation}
E=\frac{1}{ma^2}(1-\cos ka),
\label{dispersion2}
\end{equation}
thus $k(E)$ in equation (\ref{wf2}) is given by
\begin{equation}
k(E)=\arccos\left(1-Ema^2\right).
\label{invdispersion2}
\end{equation}
[Note that the leading term in the series expansion of equation (\ref{dispersion2}) is $k^2/2m,$ in accordance with (\ref{invdispersion}).]

\bigskip
Despite discretization, the problem is still infinite dimensional: the grid points $x_j$ range from $-\infty$ to $\infty.$ More specifically, the integer index $j$ has values in the interval $(-\infty,-1]$ for region I, $[0,M]$ for region II, and $[M+1,\infty)$ for region III. However, it can be shown \cite{D95,P15} that it is sufficient to concentrate on the spatial domain where the potential is nonzero, i.e., on region II. (For constant potentials, simple matrix multiplication leads to this result, but it holds also in our case.) The effect of the semi-infinite domains I and III are taken into account by so-called self energy terms (memory kernels) in the Dyson equation:
\begin{equation}
i\frac{\partial}{\partial t_1}\mathbf{G}^R(t_1,t_2)=\mathbf{H}(t_1)\mathbf{G}^R(t_1,t_2)+\int\limits_{t_2}^{t_1}  \mathbf{\Sigma}^R(t_1,t) \mathbf{G}^R(t,t_2) dt,
\label{Dyson}
\end{equation}
where boldface symbols emphasize matrix properties of the objects (the products of which are meant to be matrix multiplication). Note that due to the retardation of the functions in the integrand ($t_1>t, \ t>t_2$), the limits of the integral can be extended to $\pm\infty.$ The nonzero elements of $\mathbf{\Sigma}^R$ are related to the semi-infinite domains:
\begin{equation}
\label{SigmaR}
\Sigma^R_{i,j}(t,t')=\left\{\begin{array}{ll}
      d^2g^R_{10}(t,t')& \mathrm{if}\ i=j=1 \ \mathrm{or}\ i=j=M,\\
      0 & \mathrm{otherwise},
\end{array}
\right.
\end{equation}
where $\mathbf{g}^R(t,t')$ denotes the retarded Green's matrix of a semi-infinite discretized line segment not being coupled to the central region (II), and $d=\frac{1}{2ma^2}$ is the offdiagonal or 'hopping' matrix element of the Hamiltonian (\ref{Ham}). The index $0,1$ means that one needs the matrix element between the termination point (indexed by $0$) and the only neighbor it has.

In agreement with equation (\ref{GRzero}), the initial condition for the integro-differential equation (\ref{Dyson}) is
\begin{equation}
\lim_{t_1 \searrow t_2}G^R_{jk}\left(t_1,t_2\right)=-i \delta_{jk}.
\label{initial}
\end{equation}
Since $\mathbf{\Sigma}^R$ can be calculated analytically (see the next section), equation (\ref{Dyson}) together with (\ref{initial}) allows us -- at least in principle -- to calculate the retarded Green's function (matrix) for arbitrary two spacetime points. However, according to the traditional treatment, one also needs to calculate the lesser Green's  functions, in order to obtain physically interpretable results. In other words, the Keldysh equation
\begin{equation}
\mathbf{G}^<(t_1,t_2)=\int\limits_{-\infty}^{t_1}\int\limits_{-\infty}^{t_2}  \mathbf{G}^R(t_1,t)\mathbf{\Sigma}^<(t,t') \left[\mathbf{G}^R(t_2,t')\right]^\dagger dt dt'
\label{Keldysh}
\end{equation}
has to be solved as well. The lesser self energy $\mathbf{\Sigma}^<$ generally corresponds to scattering from regions I and III into the central one. However, for the boundary conditions that we consider (incoming monoenergetic plane wave from the left, i.e., from region I), region III has no influence. Similarly to $\mathbf{\Sigma}^R,$ the matrix $\mathbf{\Sigma}^<$ can be expressed using the lesser Green's function $\mathbf{g}^<$ of a "standalone" semi-infinite line segment:
\begin{equation}
\label{Sigmaless}
\Sigma^<_{i,j}(t,t')=\delta_{ij}\delta_{i0}\ d^2g^<_{i,j}(t,t').
\label{Sless}
\end{equation}
The analytical form of the single nonzero matrix element will be given in the next section.

\section{Solvable, numerically exact dynamical equations}
In this section, first we give explicit expressions for the nonzero matrix elements of $\mathbf{\Sigma}^R$ and $\mathbf{\Sigma}^<,$ discuss the structure of the Dyson and Keldysh equations, and finally introduce numerically exact dynamical equations that can be solved without considerable computational costs.

The key observation for the determination of the Green's functions of the semi-infinite line segments (region I: input, region III: output) is that since the Hamiltonian acting on these domains has no explicit time dependence, time translation is a symmetry of the problem. Practically, $\mathbf{g}^R(t,t')=\mathbf{g}^R(\tau)$ and $\mathbf{g}^<(t,t')=\mathbf{g}^<(\tau),$ with $\tau=t-t'.$ This allows us to perform a Fourier transformation with respect to the time difference $\tau,$ the conjugate variable of which is the energy ($\hbar=1$). The retarded Green's function of a semi-infinite line segment in energy domain can be calculated e.g., by eigenfunction expansion and contour integration \cite{D95,P15}. The result is
\begin{equation}
g^R_{10}(E)=-\frac{1}{d}e^{ik(E)a},
\end{equation}
where $a$ is the spatial step size and $k(E)$ is given by equation (\ref{invdispersion2}). That is,
\begin{equation}
g^R_{10}(E)=-\frac{1}{d}\left[1-Ema^2+i \sqrt{1-\left(1-Ema^2\right)^2} \right].
\end{equation}
The inverse Fourier transform of this expression can be calculated analytically, leading to:
\begin{equation}
g^R_{10}(\tau)=-i\Theta(\tau)\frac{J_1(2\tau d)}{\tau d}e^{-2i\tau d},
\label{gR}
\end{equation}
where first order Bessel's function of the first kind appear. Since in our case $\mathbf{\Sigma}^<(\tau)=d^2\mathbf{g}^<(\tau)$ describes monoenergetic scattering into region II, its time dependence is particularly simple, it contains a single Fourier component:
\begin{equation}
\Sigma^<_{k,l}(\tau)=i d^2 N(E) e^{-iE\tau} \delta_{kl}\delta_{k0},
\label{gless}
\end{equation}
where $E$ denotes the energy of the incoming plane wave (\ref{wf}). As we shall see later, the convention that the constant particle density that corresponds to equation (\ref{wf}) is unity, implies the normalization: $N(E)=\sqrt{1-\left(1-\frac{E}{2d}\right)^2}.$

\bigskip

Being equipped with the explicit expressions (\ref{gR}) and (\ref{gless}), we can turn to the analysis of the structure of equations (\ref{Dyson}) and (\ref{Keldysh}). First, let us note that the retarded Green's function (in our discrete case: matrix) of region II, the dynamics of which is determined by the Dyson equation (\ref{Dyson}), is general in the sense that its knowledge is sufficient to determine physical quantities corresponding to different possible boundary conditions given by $\mathbf{\Sigma}^<(\tau).$ Although in the current paper we are not going to investigate boundary conditions other than monoenergetic input from region I [that corresponds to equation (\ref{gless})], it is worth emphasizing that this is not the only possibility.

\begin{figure}[tbh]
\centering
\includegraphics[width=12cm]{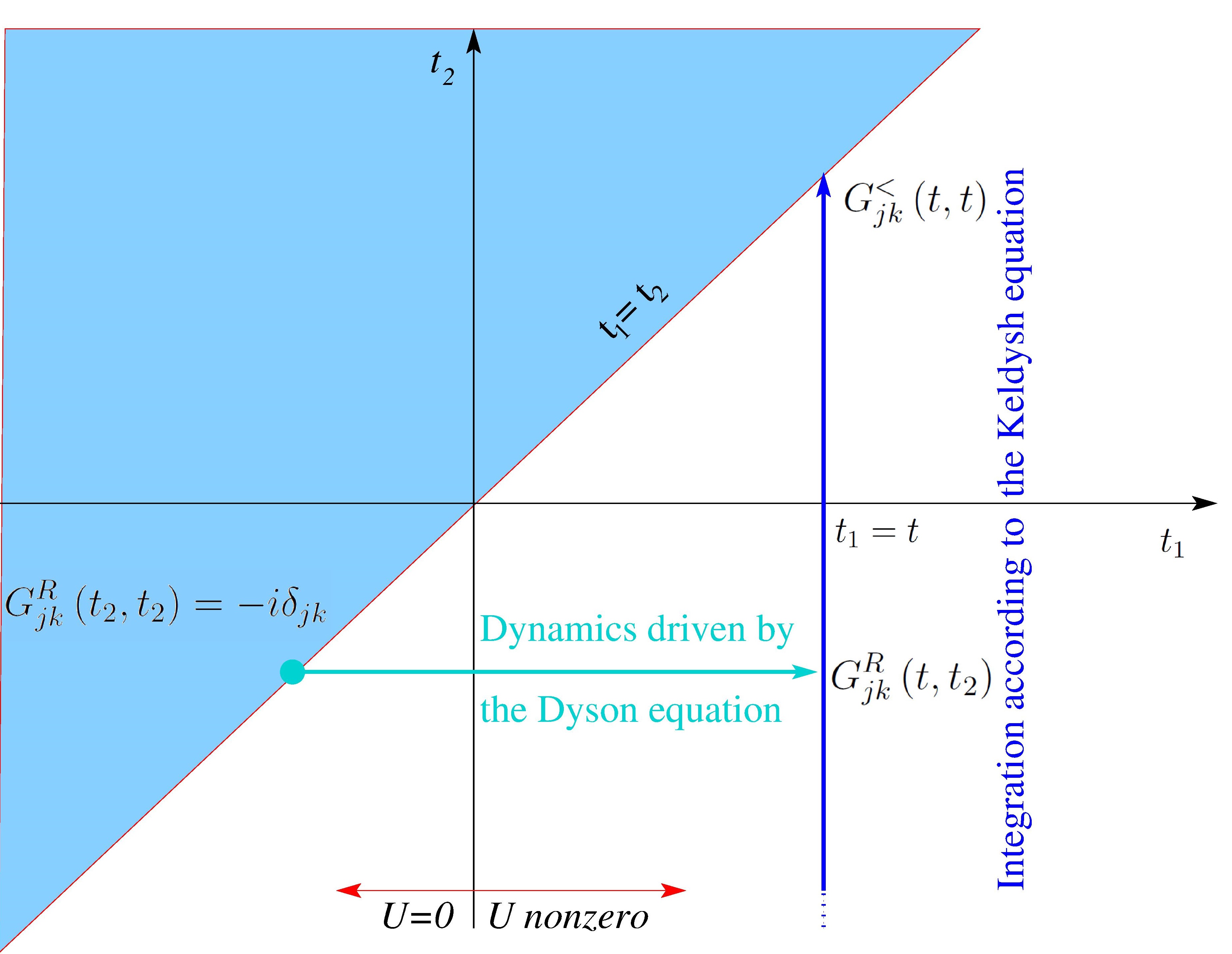}
\caption{(Color online) The structure of the equations to be solved in order to obtain the particle density that is proportional to $G^<_{ii}\left(t,t\right).$}
\label{solutionfig}
\end{figure}

The structure of the equations to be solved in order to obtain e.g. the local particle density is summarized in figure \ref{solutionfig}. Using equations (\ref{gless}) and (\ref{Sless}), it is easy to see that if we are to obtain $G^<_{ii}(t,t)$, the Keldysh equation (\ref{Keldysh}) factorizes:
\begin{equation}
G^<_{ii}(t,t)=i d^2 N(E)\left|\int\limits_{-\infty}^{t}  G^R_{i0}(t,t')d^2 e^{-iEt'}  dt' \right|^2.
\label{Keldysh2}
\end{equation}
Additionally, only the $0$th column of the matrix $\mathbf{G}^R$ is needed, and as one can check easily by matrix multiplication, the Dyson equation (\ref{Dyson}) is closed to these matrix elements (i.e., no other column than the $0$th appear on the right hand side, provided this holds also for the left hand side). That means that the problem related to a $(M+1)\times (M+1)$ matrix reduces to the dynamics of a vector of dimension $M+1.$
This means that for a given time instant $t_1,$ the time evolution for the relevant part of $\mathbf{G}^R(t_1,t_2)$ [i.e., $G_{i0}^R(t_1,t_2) \ i=0,\ldots,M$] can be calculated effectively. (Note that the integral kernel modifies the time derivative of the matrix elements $G^R_{00}$, $G^R_{M0}$ only, and e.g., usual routines for a set of ordinary differential equations work perfectly.) However, the lower limit of the integral in equation (\ref{Dyson}) still means a difficulty. As a possible way, one may use real time decomposition and omit the "irrelevant" part of the retarded Green's function as in reference \cite{Z05}, but in our case there is a more efficient approach that completely avoids numerical integration in the Keldysh equation.

The idea is calculating the time evolution of the integral appearing in equation (\ref{Keldysh2}) directly. In order to simplify the notation, let us introduce the following column vector:
\begin{equation}
\Psi_i(t)=\sqrt{N(E)}d \int\limits_{-\infty}^{t}  G^R_{i0}(t,t') e^{-iEt'}  dt'.
\label{Psi}
\end{equation}
The linearity of the Dyson equation allows us to write
\begin{equation}
i\frac{\partial}{\partial t}\mathbf{\Psi}(t)=\mathbf{H}(t)\mathbf{\Psi}(t)+\int\limits_{-\infty}^{t}  \mathbf{\Sigma}^R(t-t') \mathbf{\Psi}(t') dt'.
\label{Dyson2}
\end{equation}
It is worth separating the solution corresponding to $H^0$ [the Hamiltonian with no potential term, see equation (\ref{H0})], that is, assuming
\begin{equation}
\mathbf{\Psi}=\mathbf{\Psi}^0 +\mathbf{\Psi}^1,
\label{psisum}
\end{equation} where
\begin{equation}
\Psi^0_i(t)=\sqrt{N(E)}d \int\limits_{-\infty}^{t}  G^{R0}_{i0}(t,t') e^{-iEt'}  dt'.
\label{Psi12}
\end{equation}
Here, $G^{R0}$ obeys equation (\ref{Dyson}) with $U=0$ and the initial conditions given by (\ref{initial}). This means that $G^{R0}$ is the retarded Green's function of the entire potential-free discretized infinite line that is evaluated in region II, and consequently it is known analytically: $G_{n0}^{R0}(t,t')=i^n \Theta(t-t') J_n[2(t-t')d] \exp [-2i(t-t')d],$ where Bessel functions of the first kind appear again. Combining this and the analytic (but lengthy) expression for $\int\limits_{-\infty}^{0} J_n(t)\exp(i\omega t) dt,$ one obtains
\begin{equation}
\Psi^0_n(t)=e^{i\left[k(E)na-Et\right]},
\label{Psi0}
\end{equation}
which is just the continuation of the monochromatic plane wave that arrived from the left hand side boundary of region II. (Note the disappearance of the normalization factor.) As one can check easily, the equation of motion for $\mathbf{\Psi}^0$ is the following:
\begin{equation}
i\frac{\partial}{\partial t}\mathbf{\Psi}^0(t)=\mathbf{H}^0\mathbf{\Psi}^0(t)+\int\limits_{-\infty}^{t}  \mathbf{\Sigma}^R(t-t') \mathbf{\Psiˇ}^0(t') dt'.
\label{Dyson0}
\end{equation}

Equation (\ref{Psi0}) implies that all perturbations induced by the space and time-dependent potential $U(x,t)$ is encoded into $\mathbf{\Psi}^1.$ By substracting equation (\ref{Dyson0}) from (\ref{Dyson2}), we obtain
\begin{equation}
i\frac{\partial}{\partial t}\mathbf{\Psi}^1(t)=\mathbf{U}(t)\mathbf{\Psi}^0(t) + \mathbf{H}(t)\mathbf{\Psi}^1(t)+\int\limits_{0}^{t}  \mathbf{\Sigma}^R(t-t') \mathbf{\Psiˇ}^1(t') dt'.
\label{Dyson1}
\end{equation}
Additionally, $\mathbf{\Psiˇ}^1(t)=0$ for $t<0,$ and it remains zero, unless the potential becomes finite; the source term in equation (\ref{Dyson1}) is the first one on the right hand side.

\bigskip

The central equations that provide a solution to the scattering problem we outlined in Sec.~\ref{problemsec} are (\ref{psisum}), (\ref{Psi0}) and (\ref{Dyson1}). They allow for a clear physical interpretation, for which it is the simplest to recall equations (\ref{dens}) and (\ref{Keldysh2}) to see
\begin{equation}
n(x=ia,t)=\left|\Psi_i(t)\right|^2.
\end{equation}
Additionally, equation (\ref{Dyson2}), which governs the time evolution of $\mathbf{\Psi}(t),$ is essentially a Schr\"odinger equation, apart from points $0$ and $M,$ where the integral with memory kernel mimics the presence of the semi-infinite line segments I and III, respectively. (That is, it provides numerically exact transparent boundary conditions.) Moreover, for $U=0,$  $\mathbf{\Psiˇ}(t)=\mathbf{\Psiˇ}^0(t),$ which is the plane wave input that propagates in an unperturbed way.

In summary, the complex valued, space and time-dependent function $\mathbf{\Psi}(t)$ can simply be identified with the single particle quantum mechanical wave function.

\section{An application: short laser pulse interacting with an electron beam}
Let us consider an electron beam propagating along the $x$ axis. After $t=0,$ a short, linearly polarized laser pulse impinges on the beam from a perpendicular direction. The space and time dependence of the laser field is assumed to be
\begin{equation}
\mathbf{E}(x,t)=\left\{\begin{array}{ll}
      &\hat{\mathbf{x}}\mathcal{E}_0\cos(\omega_0
t+\varphi_{\mathrm{CEP}}) \sin^2(t\pi/T) \sin^2(x\pi/L) \\ &\mathrm{if}\ x\in [0,L] \ \mathrm{and}\ t\in [0,T],\\
\\
      &0 \  \mathrm{otherwise},
\end{array}\right.,
\label{pulse}
\end{equation}
where $T$ characterizes the duration of the pulse. For ultrashort laser pulses, $T$ corresponds to few optical cycles only and the carrier-envelope phase $\varphi_{\mathrm{CEP}}$ (that determines the "waveform" of the pulse) can play an important role \cite{K98,C07}. Note that the spatial and temporal envelopes ($\sin^2$ functions) have been chosen to ensure smooth on/off switching.

We assume that dipole approximation is valid, thus the potential term in the Hamiltonian can be written as
\begin{equation}
U(x,t)=-\int\limits_{0}^{x} x'E(x',t) dx',
\end{equation}
where we used atomic units ($e=1$). This means that $U$ is not local in the sense of the previous sections [see equation (\ref{potential})], in region III it is constant in space, but oscillates as function of time: $U_{III}(t)=U(L,t).$ Consequently, we have to modify our results to be able to apply them to this problem. Returning for a moment to the continuous notation of the space variable $x$ (and omitting boldface symbols that indicated matrix-vector objects), a possible solution is to look for the time-dependent wave function in the form of
\begin{equation}
\Psi(x,t)=u(x,t) \phi(x,t)= u(x,t)\left[\Psi^0(x,t) + \phi^1(x,t)\right],
\end{equation}
where $u(x,t)=\exp[-i\int\limits_{0}^{t} U(x,t') dt'].$ Substituting back to equation (\ref{Dyson2}), we obtain:
\begin{eqnarray}
i\frac{\partial}{\partial t}{\phi}(x,t)=&u^*(x,t)H^0u(x,t) \phi(x,t) \nonumber \\
&+ u^*(x,t)\int\limits_{-\infty}^{t}  \Sigma^R(t-t') \phi(t') u(x,t') dt'.
\label{Dysonpot}
\end{eqnarray}
Subtracting equation (\ref{Dyson0}) (that describes the time derivative of a monoenergetic plane wave) from the equation above, an equation of motion for $\phi^1(x,t)$ can be deduced:
\begin{eqnarray}
i\frac{\partial}{\partial t}{\phi^1}(x,t)=&\tilde {H}^0(t)\phi^1(x,t) + (\tilde {H}^0(t)-H^0)\Psi^0(x,t) \nonumber \\
&+u^*(x,t)\int\limits_{0}^{t}  \Sigma^R(t-t') \phi^1(t') u(x,t') dt',
\label{Dysonpot1}
\end{eqnarray}
where $\tilde {H}^0(t)=u(x,t)H^0u^*(x,t)$ and the perfectly transparent boundary property has been used to simplify the memory kernel.

\begin{figure}[tbh]
\centering
\includegraphics[width=12cm]{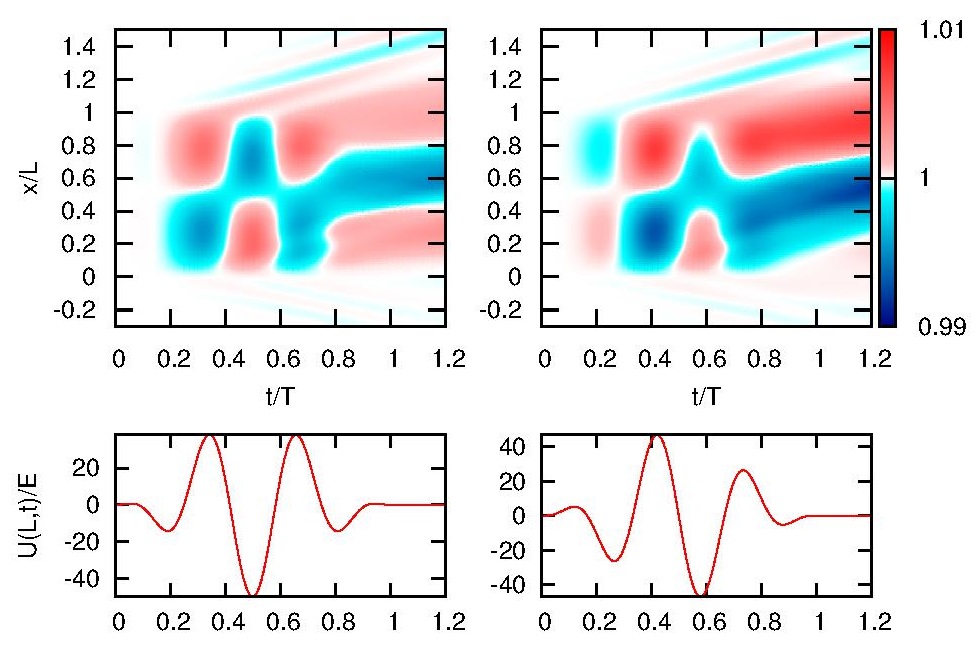}
\caption{(Color online) Laser induced density fluctuations in a monoenergetic electron beam. The wave function before $t=0$ is given by equation (\ref{wf}), and the electron density $n(x,t)$ (\ref{dens}) is shown in the top row as a contour plot. The difference between the columns is the carrier-envelope phase of the driving laser field [see equation \ref{pulse}], $\varphi_{\mathrm{CEP}}=0$ for the left column, $\varphi_{\mathrm{CEP}}=\pi/2$ for the right one. (As a reference, the time dependence of the potential $U(x,t)$ at $x=L$ is shown in the bottom row.)}
\label{resultfig}
\end{figure}

As an example, figure \ref{resultfig} shows the fluctuations of the electron density that arise due to the excitation by the laser pulse. Since -- according to our remark following equation (\ref{dens}) -- $n(x,t)=1$ corresponds to the undisturbed plane wave (\ref{wf}), we have chosen a color scheme where this value is denoted by white (i.e., invisible on the paper). As we expect, the disturbance caused by the laser field becomes visible when the potential gets significantly different from zero. The fluctuations, on the other hand, propagate away from the central region, and this wave-like behavior is still present when $t>T,$ i.e., when the laser pulse is over and the potential is globally zero again. This behavior is related to the parameters: the duration of the pulse (\ref{pulse}) has the same order of magnitude as $2\pi/E$ in figure \ref{resultfig} (concretely: ET=2), which means that the system does not follow the change of the potential adiabatically. However, the laser pulse does not mean very short, "delta-excitation" either, since, as we can see in the figure, the details of the time dependence of $U(x,t)$ strongly modify the function $n(x,t)$, the electron density fluctuations are qualitatively different for $\varphi_{\mathrm{CEP}}=0$ and $\varphi_{\mathrm{CEP}}=\pi/2:$ the system exhibits CEP dependence.

\section{Summary}
We described an efficient way of solving scattering problems with time-dependent potentials that are localized both in space and time. We have shown that the theory of time-dependent nonequlibrium Green's functions provides numerically exact dynamical equations on a spatial grid so that the boundary conditions are also taken into account appropriately. As an example, we have shown that short laser pulses (that contain a few optical cycles only) interacting with an electron beam can create electron density fluctuations the detailed structure of which depends on the parameters of the exciting pulse.

\ack
This work was partially supported
by the European Union and the European Social Fund through project
entitled "ELITeam at the University of Szeged",
and by the Hungarian Scientific Research Fund (OTKA) under Contract No.~116688.
The ELI-ALPS project (GOP-1.1.1-12/B-2012-0001) is supported by the European Union
and co-financed by the European Regional Development Fund.

\section*{References}

\providecommand{\newblock}{}

\end{document}